\documentclass[a4paper,10pt]{article}
\usepackage[dvips]{graphicx}
\usepackage{amssymb,amsmath}
\numberwithin{equation}{section}

\begin{document}

\title{Accelerating cosmology in F(T)  gravity with scalar field }
\author{K.K.Yerzhanov, Sh.R.Myrzakul, I.I.Kulnazarov, R.Myrzakulov\footnote{The corresponding author. Email: cnlpmyra1954@yahoo.com} \\ \textit{Eurasian International Center for Theoretical Physics,} \\ \textit{Eurasian National University, Astana, 010008, Kazakhstan}}

\date{}

\maketitle

\begin{abstract}
This work deals with $F(T)$ gravity models driven by real scalar fields with usual and phantom dynamics. We illustrate the results with examples of current interest, and we find some analytical solutions for scale factors and scalar fields. The results indicate that torsion-scalar models also admit the accelerated expansion of the universe. 
\end{abstract}
\vspace{2cm} \sloppy


\section{Introduction}

Various cosmological observations (the Type Ia Supernova, the cosmic microwave background radiation, the large scale structure, and so on) have shown that the universe is undergoing an accelerating expansion and it entered this accelerating phase only in the near past (see, e.g. \cite{L1} for recent  review). This unexpected discovery of the accelerated expansion of the universe  has opened one of the most puzzling and deepest problems in cosmology today. 
 Although the cosmological constant seems to be the simplest explanation for the phenomenon, several theoretical models (such as  quintessence, phantom, quintom, the (generalized) Chaplygin gas, and so on) have been presented. While some authors sustain the idea of the existence of a dark energy, others propose modifications of the General Relativity  Lagrangian known as $F(R)$
(see, e.g. \cite{N1}-\cite{EMOG} and references therein) as a way to obtain a late accelerating expansion. A great diffculty these theories have, from the point of view of the metric formalism, is that the resulting field equations are 4th order equations, which in many cases makes these hard to analyze.  Alternatively, the Palatini variational approach for such $F(R)$ theories leads to 2nd order field equations, and some authors have achieved to put observational constraints to these theories. However in many cases the equations are still hard to work with, as evidenced by the functional form of the modified Friedmann equation for a generic $F(R)$. 

Recently, models based on modified teleparallel gravity   were presented as an alternative to inflationary
 models \cite{FF1}-\cite{M1}. The theory so obtained is called as the $F(T)$ theory (see e.g. \cite{M1}).
In this theory instead  the curvature defined via the Levi-Civita connection,  the so-called Weitzenb$\ddot{o}$ck connection is used \cite{Wei}. But in this case the theory has no curvature but instead torsion.  Similar to general relativity where the action is the curvature scalar, the action of teleparallel gravity is a torsion scalar $T$. In analogy to the $F(R)$ theory,  in Refs. \cite{FF1}-\cite{M1} by generalizing the action of teleparallel gravity and found that it can explain the observed acceleration of the universe. Let us also note here that models based on modified teleparallel gravity may also provide an alternative to inflation \cite{FF1}-\cite{FF2}. Another advantage the generalized $F(T)$ torsion theory has is that its field equations are second order as opposed to the fourth order equations of
$F(R)$ theory.   In \cite{WuYu} were  presented  the constraints on some  models of the $F(T)$ gravity from the latest observational data, including the Type Ia supernovae released by the Supernova Cosmology Project collaboration, the baryonic acoustic oscillation from the spectroscopic Sloan Digital Sky Survey and the cosmic microwave background radiation from Wilkinson Microwave Anisotropy Probe seven year observation.

In this paper, we investigate extensions of the $F(T)$ gravity  models where the torsion will be the responsible of the observed acceleration of the universe but additionally driven by real scalar fields with usual and phantom dynamics. We illustrate the results with examples of current interest, and we find some analytical solutions for scale factors and scalar fields. The results indicate that torsion-scalar models also admit the accelerated expansion of the universe.

\section{$F(T)$ gravity}
Let us we start with the following action for the $F(T)$ gravity \cite{L2}-\cite{BF}
\begin {equation}
S=\frac{1}{2k^2}\int dx^{4}[\sqrt{-g}F(T)+L_{m}],
\end{equation}
where $T$ is the torsion scalar, $F(T)$ is general differentiable function of the torsion and $L_{m}$ corresponds to the matter Lagrangian, $k^2=8\pi G$. Here the torsion scalar $T$ is defined as (see e.g. \cite{HS1}-\cite{BF})
\begin{align}
	T=S_\rho^{\mu\nu}T^\rho_{\mu\nu},
\end{align}
where
\begin{align}
	S_\rho\,^{\mu\nu}=\frac{1}{2}(K^{\mu\nu}\,_\rho+\delta^\mu_\rho T^{\theta\nu}\,_\theta-\delta^\nu_\rho T^{\theta\mu}\,_\theta).
\end{align}
Here  $K^{\mu\nu}_{\rho}$ is the contorsion tensor:
\begin{align}
K^{\mu\nu}\,_\rho=-\frac{1}{2}(T^{\mu\nu}\,_\rho-T^{\nu\mu}\,_\rho-T_\rho\,^{\mu\nu}),
\end{align}
which equals the difference between Weitzenb$\ddot{o}$ck and Levi-Civita connections. So in $F(T)$ gravity uses the curvatureless Weitzenb$\ddot{o}$ck connection, whose non-null torsion is
\begin{align}
T^\lambda_{\mu\nu}=\stackrel{w}{\Gamma}^\lambda_{\nu\mu}-\stackrel{w}{\Gamma}^\lambda_{\mu\nu}=e^\lambda_i(\partial_\mu e^i_\nu-\partial_\nu e^i_\mu).
\end{align}
Here take place the following formulas
\begin{align}
g_{\mu\nu}(x)=\eta_{ij}e^i_\mu(x)e^j_\nu(x), 
\end{align}
where
\begin{align}
{\bf e}_i \cdot {\bf e}_j=\eta_{ij}, \quad  \eta_{ij}=diag(1, -1, -1, -1).
\end{align}
The varition of  the action (2.1) with respect  to the vierbein gives the following field equations \cite{BF}
\begin {equation}
[e^{-1}\partial_{\mu}(eS^{\mu\nu}_{i}-e^{\lambda}_{i}T^{\rho}_{\mu\lambda}S^{\nu\mu}_{\rho}]F_{T}+S^{\mu\nu}_{i}\partial_{\mu}TF_{TT}+\frac{1}{4}e^{\nu}_{i}F=\frac{1}{2}k^2e^{\rho}_{i}T^{\nu}_{\mu}.
\end{equation}
Here $e=\sqrt{-g}$, $S^{\mu\nu}_{i}=e^{\rho}_{i}S^{\mu\nu}_{\rho}$ and $T_{\mu\nu}$ is the matter energy-momentum tensor.  As we see, Eq.(2.8) are 2nd order makes them simpler than the corresponding field equations resulting in the other modified gravity theories like $F(R), F(G)$ and so on.   We now will assume a flat homogeneous and isotropic FRW universe with  the metric 
\begin {equation}
ds^{2}=-dt^{2}+a(t)^{2}\sum^{3}_{i=1}(dx^{i})^{2},
\end{equation}
where $t$ is  cosmic time. For this metric  
\begin{align}
	e_\mu^i=diag(1, a(t), a(t), a(t)), \quad H=\frac{\dot{a}}{a}, \quad T=-6H^2
\end{align}
and   the modified Friedmann equations and the conservation equation  are given by
\begin{align}
	12H^2 F_{T}+F=2k^2 \rho, 
\end{align}
\begin{align}
	48 H^2 \dot{H}F_{TT}-(12H^2+4\dot{H})F_{T}-F=2k^2p, 
\end{align}
\begin{align}
	\dot{\rho}+3H(\rho+p)=0.
\end{align}
This set of equations we can rewrite as
\begin{align}
	-2TF_{T}+F=2k^2 \rho, 
\end{align}
\begin{align}
	-8\dot{H}TF_{TT}+(2T-4\dot{H})F_{T}-F=2k^2p, 
\end{align}
\begin{align}
	\dot{\rho}+3H(\rho+p)=0.
\end{align}
 Note that the last equation can be written as  \cite{BF} 
\begin{align}
	\frac{d}{dt}(a^3\rho)=-3a^3Hp.
\end{align}
Also we note that as $F(T)=T$ Eqs. (2.14)-(2.15) transform to the usual Friedmann equations of GR
\begin{equation}
\frac{3}{k^2}H^2=\rho, \quad \frac{1}{k^2}(2\dot{H}+3H^2)=-p.
\end{equation}
So we can rewrite the equations (2.14)-(2.15) as
\begin{equation}
\frac{3}{k^2}H^2=\rho+\rho_T, \quad \frac{1}{k^2}(2\dot{H}+3H^2)=-(p+p_T),
\end{equation}
where
\begin{equation}
\rho_T=\frac{1}{2k^2}(2TF_T-F+6H^2), \quad p_T=-\frac{1}{2k^2}[-8\dot{H}TF_{TT}+(2T-4\dot{H})F_{T}-F+4\dot{H}+6H^2]
\end{equation}
are the torsion contributions to the energy density and pressure. We also present the parameter of state  and the deceleration parameter
\begin{align}
	w=-1+\frac{-8\dot{H}TF_{TT}-4\dot{H}F_{T}}{-2TF_{T}+F}, \quad q=-\frac{\ddot{a}}{aH^2},
\end{align}
respectively.
\section{The torsion-scalar model}
Let us we now introduce two functions $\phi, V$ as
\begin{align}
\epsilon\dot{\phi}^2=-8\dot{H}TF_{TT}-4\dot{H}F_{T}, \quad V=4\dot{H}TF_{TT}-2(T-\dot{H})F_{T}+F,
\end{align}
where dot denotes the time derivative. Hence we get 
\begin{align}
	-2TF_{T}+F=2k^2 [\frac{1}{2}\epsilon\dot{\phi}^2+V], 
\end{align}
\begin{align}
	-8\dot{H}TF_{TT}+(2T-4\dot{H})F_{T}-F=2k^2[\frac{1}{2}\epsilon\dot{\phi}^2-V], 
\end{align}
\begin{align}
	\ddot{\phi}+3H\dot{\phi}+\epsilon\frac{\partial V}{\partial\phi}=0,
\end{align}
where $\epsilon=1$ for the usual case and $\epsilon=-1$ for the phantom case. If we compare these equations with (2.14)-(2.16) we see that
\begin{align}
	\rho=\frac{1}{2}\epsilon\dot{\phi}^2+V, \quad p=\frac{1}{2}\epsilon\dot{\phi}^2-V. 
\end{align}
From here we restrict ourselves for simplicity to the case   $F=\alpha T+\beta T^{0.5}$. Then
\begin{align}
\epsilon\dot{\phi}^2=-4\alpha\dot{H}, \quad V=-\alpha T+2\alpha\dot{H}.
\end{align}
and 
\begin{align}
w=-1+4\frac{\dot{H}}{T}.
\end{align}
Now we consider some examples.
\section{Reconstruction for a given $a(t)$}
\subsection{Case I: $a=\delta \sinh^{m}[\mu t]$}
Our first example is 
\begin{align}a=\delta \sinh^{m}[\mu t].
\end{align}
 For this case we have
\begin{align}
H=\mu m\coth[\mu t], \quad \dot{H}=-\frac{\mu^2m}{\sinh^2[\mu t]},\quad \dot{\phi}^2=\frac{4\alpha\mu^2m}{\epsilon \sinh^2[\mu t]}.
\end{align}
Hence we get
\begin{align}
\phi=\phi_0\pm2\sqrt{\frac{\alpha\mu^2m}{\epsilon}}\log[\tanh[\frac{\mu t}{2}]],\quad V=6\alpha m^2 \mu^2\coth^2[\mu t]-\frac{2\alpha\mu^2m}{ \sinh^2[\mu t]}.
\end{align}
So the potential takes the form ( $\tanh[\frac{\mu t}{2}]=e^{\pm\frac{\phi-\phi_0}{2\sqrt{\alpha\mu^2 m \epsilon^{-1}}}}$)
\begin{align}
V=3\alpha m^2 \mu^2[\frac{1+e^{\pm\frac{\phi-\phi_0}{\sqrt{\alpha\mu^2 m \epsilon^{-1}}}}}{e^{\pm\frac{\phi-\phi_0}{2\sqrt{\alpha\mu^2 m \epsilon^{-1}}}}}]-
\frac{\alpha\mu^2m(1-e^{\pm\frac{\phi-\phi_0}{\sqrt{\alpha\mu^2 m \epsilon^{-1}}}})^2}{2e^{\pm\frac{\phi-\phi_0}{\sqrt{\alpha\mu^2 m \epsilon^{-1}}}}}.
\end{align}

\subsection{Case II: $a=a_0e^{\beta t^m}$} 

Let  $a=a_0e^{\frac{\delta}{m+1} t^{m+1}}$. Then  $H=\delta t^m$ and we have  
\begin{align}
t=[\frac{(\phi-\phi_0)(m+1)}{\pm 4\sqrt{-\alpha m\delta\epsilon^{-1}}}]^{\frac{2}{m+1}},\quad \epsilon\dot{\phi}^2=-4\alpha m\delta t^{m-1},
\end{align}
and hence
\begin{align}
\phi=\phi_0\pm\frac{4\sqrt{-\alpha m\delta\epsilon^{-1}}}{m+1}t^{\frac{m+1}{2}}, \quad V=6\alpha\delta^2 t^{2m}+2\alpha m\delta t^{m-1}.
\end{align}
Then finally we get
\begin{align}
V=6\alpha\delta^2[\frac{(\phi-\phi_0)(m+1)}{\pm4\sqrt{-\alpha m\delta\epsilon^{-1}}}]^{\frac{4m}{m+1}}+2\alpha m \delta [\frac{(\phi-\phi_0)(m+1)}{\pm 4\sqrt{-\alpha m\delta}}]^{\frac{2(m-1)}{m+1}}.
\end{align}
\subsection{Case III: $a=a_0t^n$}
Our next  example is given by
\begin{align}
a=a_0t^n.
\end{align}
In this case we get
\begin{align}
H=\frac{n}{t}, \quad \dot{H}=-\frac{n}{t^2},\quad \epsilon\dot{\phi}^2=\frac{4\alpha n}{t^2}, \quad \phi-\phi_0=\pm 2\sqrt{\alpha n \epsilon^{-1}}\ln[t],\quad t=e^{\pm\frac{\phi-\phi_0}{2\sqrt{\alpha n \epsilon^{-1}}}}
\end{align}
and
\begin{align}
V=2\alpha n(3n-1)t^{-2}.
\end{align}
Finally we get the following expression for the potential
\begin{align}
V=2\alpha n(3n-1)e^{\mp\frac{\phi-\phi_0}{\sqrt{\alpha n\epsilon^{-1}}}}.
\end{align}
\section{Reconstruction for a given $\phi(t)$}

\subsection{Example I: $\phi=\delta\tanh[\lambda t]$}

For the case  we get the following formulas  
\begin{align}
H=H_0-0.25\alpha^{-1}\delta^2\lambda\epsilon(\tanh[\lambda t]-3^{-1}\tanh^3[\lambda t]),
\end{align}
\begin{align}
V=6\alpha[H_0-0.25\alpha^{-1}\delta^2\epsilon\lambda(\tanh[\lambda t]-3^{-1}\tanh^3[\lambda t])]^2-0.5\epsilon\delta^2\lambda^2(1-\tanh^2[\lambda t])^2.
\end{align}
In terms of $\phi$ these formulas become
\begin{align}
H=H_0-0.25\alpha^{-1}\delta^2\epsilon\lambda\phi+
12^{-1}\alpha^{-1}\delta^{-1}\lambda\epsilon\phi^3,
\end{align}
\begin{align}
V=6\alpha[H_0-0.25\alpha^{-1}\delta^2\epsilon\lambda\phi+
12^{-1}\alpha^{-1}\delta^{-1}\lambda\epsilon\phi^3]^2-0.5\epsilon\delta^2\lambda^2(1-\delta^{-2}\phi^2)^2.
\end{align}
\subsection{Example II: $\phi=\delta\cosh^{-2}[\lambda t]$}

For the case  we get the following formulas  
\begin{align}
H=H_0-\alpha^{-1}\delta^2\lambda\epsilon(3^{-1}\tanh^3[\lambda t]-5^{-1}\tanh^5[\lambda t]),
\end{align}
\begin{align}
V=6\alpha[H_0-\alpha^{-1}\delta^2\lambda\epsilon(3^{-1}\tanh^3[\lambda t]-5^{-1}\tanh^5[\lambda t])]^2-2\epsilon\delta^2\lambda^2[(1-\tanh^2[\lambda t])^2-(1-\tanh^2[\lambda t])^3].
\end{align}
In terms of $\phi$ these formulas become
\begin{align}
H=H_0-\alpha^{-1}\delta^2\lambda\epsilon[3^{-1}(1-\delta^{-1}\phi)^{\frac{3}{2}}-5^{-1}(1-\delta^{-1}\phi)^{\frac{5}{2}}],
\end{align}
\begin{align}
V=6\alpha[H_0-\alpha^{-1}\delta^2\lambda\epsilon(3^{-1}\tanh^3[\lambda t]-5^{-1}\tanh^5[\lambda t])]^2-2\epsilon\delta^2\lambda^2[(1-\tanh^2[\lambda t])^2-(1-\tanh^2[\lambda t])^3].
\end{align}

\section{Deceleration/acceleration phases. Conclusion}

In this section we would like shortly to explore the deceleration-acceleration phases in some above considered torsion-scalar models. As an example, let us consider  Case II. For this case we have
\begin{align}
w=-1-\frac{2m}{3\delta t^{m+1}},
\end{align}
\begin{align}
q=-1-\frac{m}{\delta t^{m+1}}.
\end{align}
The boundary of the deceleration and acceleration phases is  $t=t_0=(-\frac{m}{\delta})^{\frac{1}{m+1}}$ and here $q=0$ and $w=-\frac{1}{3}$. To the acceleration (deceleration) phase corresponds $t>t_0, \quad q<0, \quad w<-\frac{1}{3}$ $ (t<t_0, \quad q>0, \quad w>-\frac{1}{3})$. If $m=-1$, then these formulas take the form $\delta>1, \quad q<0, \quad w<-\frac{1}{3}$ $ (\delta<1, \quad q>0, \quad w>-\frac{1}{3})$, respectively. To compare, we note that in General Relativity an accelerated expansion is only possible if $w<-\frac{1}{3}$ that is the pressure $p=-\frac{1}{3}$ is 
negative.

Our next example is Case I for which we get
\begin{align}
w=-1+\frac{2}{3m\cosh^2[\mu  t]}, \quad
q=-1+\frac{1}{m\cosh^2[\mu  t]}.
\end{align}
The acceleration phase corresponds to $m^{-1}<\cosh^2[\mu t], \quad w<-\frac{1}{3}$ and vice versa for the deceleration phase.

In this work we have investigated generalized $F(T)$ modified torsion models, that is, models in which the torsion gravity  equations is extended to include scalar fields.  This study is a continuation of our program to investigate the $F(T)$  gravity \cite{M1}. Finally we would like to note that although to the GR corresponds the model $F(T)=T$ but our particular model $F(T)=\alpha T+\beta T^{0.5}$ also gives the same results.

\end{document}